\documentclass[usenatbib]{openjournal}
\usepackage{newtxtext,newtxmath}
\usepackage{graphicx} 
\usepackage{natbib}
\usepackage{float}
\usepackage[colorlinks,linkcolor=blue,citecolor=blue,urlcolor=blue ]{hyperref}
\begin{document}
\title{Using A One-Class SVM to Optimize Transit Detection}
\author{Jakob Roche} \address{University of South Florida}
\begin{abstract}
    As machine learning algorithms become increasingly accessible, a growing number of organizations and researchers are using these technologies to automate the process of exoplanet detection. These mainly utilize Convolutional Neural Networks (CNNs) to detect periodic dips in lightcurve data. While having approximately 5\% lower accuracy than CNNs, the results of this study show that One-Class Support Vector Machines (SVMs) can be fitted to data up to 84 times faster than simple CNNs and make predictions over 3 times faster on the same datasets using the same hardware. In addition, One-Class SVMs can be run smoothly on unspecialized hardware, removing the need for Graphics Processing Unit (GPU) usage. In cases where time and processing power are valuable resources, One-Class SVMs are able to minimize time spent on transit detection tasks while maximizing performance and efficiency.
\end{abstract}
\maketitle

\section{Introduction}
    Detecting exoplanet transit events in stellar lightcurves is a task that is increasingly becoming automated using machine learning methods. The detection of these transits, which occur when an exoplanet passes in front of its host star in relation to our line of sight, are crucial for identifying and studying exoplanets. One of the first methods used to perform machine learning based analysis on these lightcurves uses Convolutional Neural Networks (CNNs) (\cite{Pearson2017} \cite{Han2023}). These have proven effective in identifying transits, performing to a high degree of accuracy with adequate training (\cite{Priyadarshini2021}). However, CNNs are computationally demanding and require extensive manual labelling of lightcurve data.
    
    To train a CNN on lightcurve data, the data is often preprocessed by eliminating long-term trends and variations, usually due to instrumental noise or stellar variability (a process known as flattening), and removing outlying datapoints (\cite{Shallue2018}). After preprocessing, lightcurves are manually labelled as either containing a transit or not. These labelled lightcurves are fed to the CNN model, which learns to distinguish between lightcurves that contain transits and those that do not based on patterns it detects. After this training period, the model can then be used to make predictions on new, unseen lightcurve data. Since lightcurves must be manually labelled for the algorithm to function, CNNs are an example of a supervised learning algorithm (\cite{Bhatt2021}).
    
    The method described in this paper uses a One-Class Support Vector Machine (SVM) to analyze lightcurve data. This algorithm only needs one type of data (either positive or negative), unlike CNNs and other supervised learning models that need both. Because of this, this method is considered an example of semi-supervised learning, as opposed to unsupervised learning, which would require no labelled data. While CNNs are widely used to detect exoplanets, the application of One-Class SVMs for this purpose remains largely unexplored. Recent surveys of machine-learning driven exoplanet detection techniques, such as the ones by \cite{Pearson2017}, \cite{Shallue2018}, \cite{Schanche2018}, and \cite{Malik2021}, focus on CNNs and other binary classifiers, with no mention of One-Class SVMs.
    
    This paper will discuss the methodology of implementing a One-Class SVM for lightcurve analysis and compare it to more traditional CNN-based methods. Section 2 will cover data processing techniques used to reduce noise and improve signal strength in the dataset, Section 3 will discuss the specific algorithm used to perform data analysis, and Section 4 will discuss the results of this approach compared to a CNN-based approach.
\section{Data Processing}  
    To begin training the model, exoplanet-containing lightcurve data from the nine-year Kepler mission was obtained from the Mikulski Archive for Space Telescopes (MAST). Because lightcurve data that does not contain exoplanets could not be reliably sourced, exoplanet-containing data was used instead. The data was provided in the form of Pre-Search Data Conditioned Simple Aperture Photometry flux (PDCSAP flux). Pre-search data conditioning is a method used to decrease instrumental noise and other errors present in Kepler's simple aperture photometry measurements (\cite{Smith2012}).
    
    \begin{figure}[H]
        \centering
        \includegraphics[width=0.75\linewidth]{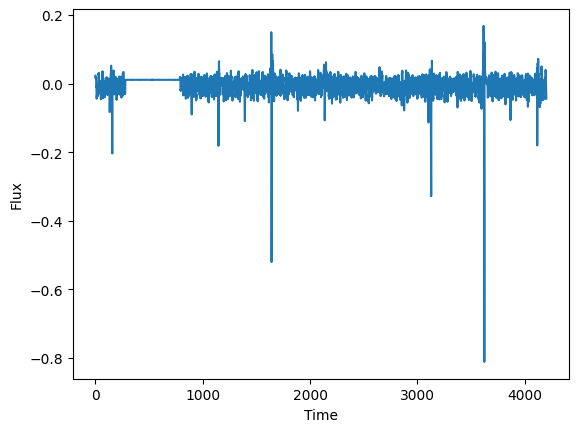}
        \caption{A PDCSAP lightcurve of Kepler-413 from the Kepler mission obtained from MAST, further cleaned and processed to reduce noise and amplify signals. This, along with 2,588 other lightcurves, comprised the training dataset. The transit of Kepler-413b, the exoplanet that orbits this star, can be seen in the smaller dips along the lightcurve. The larger drops in flux are due to an eclipsing binary in this system.}
        \label{fig:ExLightcurve}
    \end{figure}
      
    Further processing of the lightcurve data is crucial for ensuring prediction accuracy and efficiency. To improve its signal-to-noise ratio, a number of transformations were performed on the data. The PDCSAP flux was first flattened to remove any long-term stellar variability or other long-period recurring noise from the signal. This flattening was performed using a Savitzky-Golay filter (\cite{Savitzky1964}). At this point, the data took the form of a three-dimensional array with a column for the flux data as well as the time data. The end result of these transformations performed are visible in Fig. \ref{fig:ExLightcurve} which strongly preserves the transit dips while having relatively little noise. (Note that eclipsing binary signals, like the ones in Fig \ref{fig:ExLightcurve}, could not be reliably filtered out without destroying certain transit signals.) The data was then normalized using the standard scaling method to provide a common scale for all the data.
  
    The time intervals were then removed from the dataset, leaving only the processed flux in the array. This was done, in part, to avoid having the model learn, and make predictions based on, the time data. If the time data had been left in, it might have led to a scenario where the model found a pattern common to all the time data in the transit-containing dataset. Without information on what a transit-free lightcurve looks like, the model might incorrectly assume that the time data significantly influences the determination of whether the data contains a transit. In addition, removing the time data halved the size of the dataset, improving training performance.

    The flux measurements for each lightcurve were processed into rows stacked on top of each other. The finalized training dataset was a 2-dimensional array composed of 2,589 lightcurve files.

    Overall, these preprocessing steps were found to slightly increase training speed, taking, on average, 0.4 seconds off of the baseline One-Class SVM training times. However, not performing these preprocessing steps removed the model's ability to generalize (that is, make correct predictions on new data), with all the model's predictions, no matter the dataset, resulting in a 'no transit' output.
\section{Algorithms}
    \subsection{Overview of One-Class SVMs}
\begin{figure}[H]
        \centering
        \includegraphics[width=0.75\linewidth]{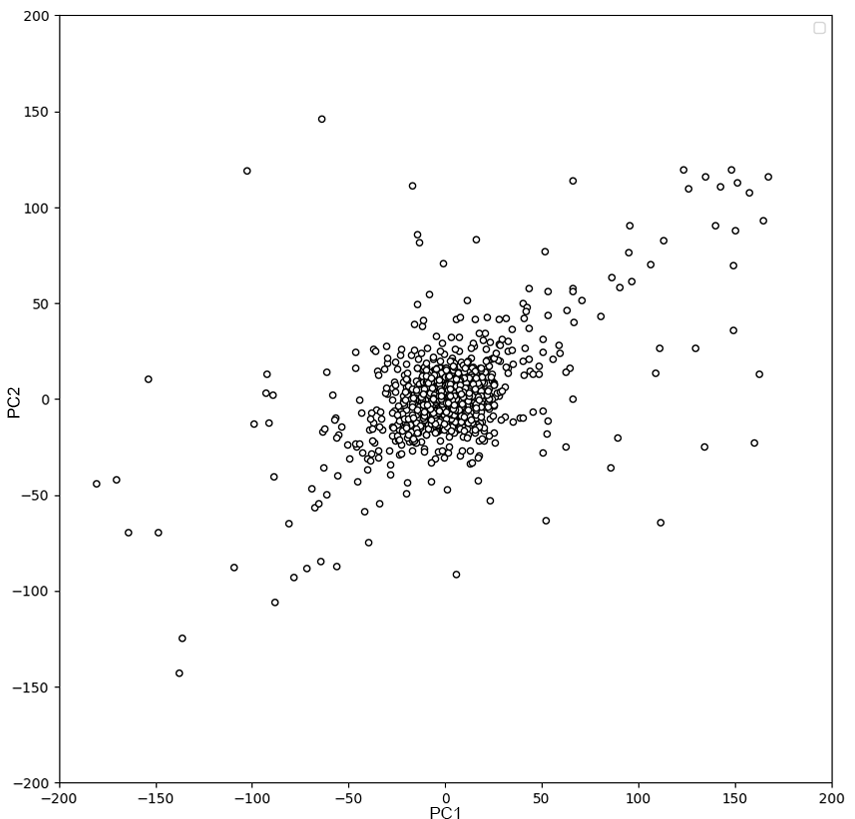}
        \caption{An enlarged 2-dimensional representation of this model's feature space. Each point is one lightcurve in the training dataset. Datapoints that are closer to each other are generally considered to be more similar than those that are further apart. Note the dense concentration of datapoints toward the center of the graph, indicating a high degree of similarity between those points. Principal component analysis was used to lower the dimensionality.}
        \label{fig:FeatureSpace}
    \end{figure}

    At its core, a One-Class SVM is an anomaly detection algorithm. This algorithm maps the datapoints (in this context, one lightcurve is equivalent to one datapoint) to specific points in a virtual space (known as feature space) based on the datapoint's specific attributes. The specific feature space generated by this model is visible in Fig. \ref{fig:FeatureSpace}. Once this feature space is created, the model then attempts to generate a virtual 'sphere' to encompass the most amount of datapoints possible. The boundary that this sphere creates between the points inside and the points outside of it is known as the decision boundary. The optimization problem (that is, the specific goal that a model has during training) of a One-Class SVM relies on finding the smallest possible radius of this sphere while still encompassing the most amount of datapoints possible (\cite{Moya1996}). Because of this, the decision boundary becomes more accurate as more data is supplied. The way the algorithm performs the optimization problem is based on specific parameters set at the beginning of model training.

    Once the correct sphere size is found, the training period ends and the model can be used to perform predictions on new data. When the model is deployed on new data, it maps the new datapoints to the same feature space and observes where they lie compared to the decision boundary. If they lie within the boundary, they are seen as being similar to the training set, and determined to be non-anomalous. If they lie outside the decision boundary, they are seen as anomalous compared to the training set.
    
    \subsection{Model Parameterization}
    
    While the basic structure of the One-Class SVM is linear, meaning that the decision boundary is essentially flat (like the surface of a sphere), more complex versions of the algorithm can use different 'kernels' to modify the shape of the decision boundary, enabling the model to learn, and make predictions on, more complex, non-linearly separable data (\cite{Bounsiar2014}). In this method, a Gaussian kernel was used.

    The Gaussian kernel measures the distance between datapoints in feature space to compare their relative similarity, with similarity decreasing with distance. This decrease follows a Gaussian distribution. It uses this similarity metric to compute a nonlinear decision boundary. The input parameter \(\gamma\) is used in the Gaussian kernel to determine the effect that a singular datapoint has on the decision boundary. This directly translates into the smoothness of the decision boundary, with a higher value of \(\gamma\) leading to a more complex, nuanced decision boundary, while a lower value leads to a smoother decision boundary (\cite{AlMejibli2020}). When \(\gamma\) is set to an extreme value in either direction, it can lead to overfitting (where a model over-trains and only recognizes datapoints in the training dataset as being correct) or underfitting (where a model does not train itself thoroughly enough). Either extreme harms the model's ability to generalize , so it's particularly important to find a suitable value for \(\gamma\).

    The value of \(\gamma\) was calculated automatically based on the following formula.
    
    \[\gamma = \frac{1}{\text{n\_features} {\times} \text{V(X)}}\]

    Where n\_features is the number of features (a feature being a measurable property of the data, in this case one flux measurement), and V(X) being the variance of the feature values. Variance, in this case, being a measure of the average dispersion or spread of the features. Using this formula helps ensure that the value of \(\gamma\) is neither too large or too small, ensuring that the highest possible accuracy is achieved.

    The second useful parameter fed to the algorithm is \(\nu\). This parameter helps control the amount of support vectors in the model. Support vectors are the datapoints closest to the decision boundary on either side of it (\cite{Hearst1998}). By modifying the value of \(\nu\), one effectively modifies the shape of the decision boundary by controlling the number of support vectors allowed. A higher \(\nu\) value leads to a more flexible model, with a more lenient boundary, whereas a lower \(\nu\) value leads to a more strict boundary. Depending on the context of the problem, both of these options can have their own benefits, but for this application a relatively low \(\nu\) value of 0.05 performed the best. In practice, this helped ensure model accuracy by making sure that lightcurves classified as containing a transit must have features very similar to the lightcurves in the training dataset. This helps limit the amount of phenomena like variable stars or other non-transit related variability misclassified by the model as being a transit.
    
    \subsection{Model Specifications}
    
    The following presents a condensed version of the model specifications used.
    
    The CNN was implemented using the Python version of the TensorFlow library, while the SVM was implemented using the Python library scikit-learn. The CNN model was designed with performance and efficiency in mind, with the amount of layers limited. In the context of a CNN, one layer is a piece of the model architecture that performs a given task. More complex models typically have more layers (\cite{Li2022}). For this setup, 13 layers were used. 
    
    The specific layers were as follows: 3 1D Convolutional Layers, 4 Dropout Layers, 3 Pooling Layers, and 3 Fully Connected Layers using Sigmoid activation. For more information on these specific layers, refer to \cite{Li2022}.

    The SVM model's source code is available at \nolinkurl{https://github.com/JakobRoche/OCSVM-Transit-Detection/}. The SVM used a Gaussian kernel with a \(\nu\) value of 0.05 and a \(\gamma\) value that was automatically calculated as specified above.

    After a period of testing and evaluation, these parameters were found to provide the model with the most amount of accuracy while still remaining computationally efficient.
    
    Once the parameters were set, the model was fit to the training dataset. After this brief period of training, the model was saved and used to run predictions on new data. The Results section discusses the numerical results of this process compared to an analogous process run with a CNN.
\section{Results}
    The One-Class SVM algorithm was chosen for this task because of its high performance and low amount of manual input needed compared to other models. As a semi-supervised learning model, it does not require as much data preparation as supervised learning models. At the same time, One-Class SVMs are much less complex than most other machine learning models, which have more computationally demanding optimization problems.

    To provide comparison, a CNN was also set up and trained on the same hardware that was used for the One-Class SVM. To ensure comparability, the CNN training dataset contained the same number of lightcurve files as the SVM training dataset. In addition, both models made predictions on the same set. The specifications of the setup are as follows.
    
    \begin{table}[H]
        \centering
        \begin{tabular}{cc}
            Operating System & Ubuntu \\
            CPU & 1 Core @ 2.20 GHz \\
            RAM & 12.67 GB \\
            Disk & 107.72 GB \\
            Training Dataset & 2,589 Files\\
            Test Dataset & 1,864 Files\\
        \end{tabular}
        \label{tab:my_label}
    \end{table}

    Note the lack of a GPU in this setup. Most modern hardware setups for machine learning utilize a GPU for its ability to perform many calculations simultaneously, which greatly accelerates the training performance of machine learning models (\cite{Amaris2016}). Every piece of hardware used is widely commercially available.

    The CNN underwent a training period of 10 epochs (one epoch being one complete dataset analyzation by the model) in an effort to make model training as efficient as possible given the lack of a GPU accelerator. Other efforts, such as limiting the amount of model layers, were made in order to allow the training process to work on a CPU.

    \begin{figure}[H]
        \centering
        \includegraphics[width=0.75\linewidth]{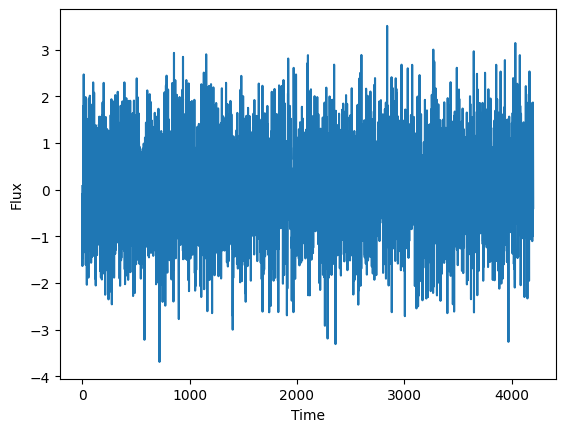}
        \caption{An artificial non-transit containing lightcurve generated to test the model's performance. The same preprocessing steps applied to the training dataset were applied to this lightcurve.}
        \label{fig:ArtificialLightcurve}
    \end{figure}

    In this context, accuracy is defined as the number of correct predictions (true positives added to true negatives) divided by the total number of predictions made. Recall is a metric that is used to find the amount of true positives correctly identified as such. It is calculated using the following formula.
    \[\text{Recall} = \frac{\text{True Positives}}{\text{True Positives}{\text{+}} \text{False Negatives}}\]
    The precision metric is used to find the amount of positive predictions that were actually correct. It is calculated using the following formula.
    \[\text{Precision} = \frac{\text{True Positives}}{\text{True Positives}{\text{+}} \text{False Positives}}\]
    
    To calculate the model's F1 score (the harmonic mean of precision and recall), a set of 1,000 artificial flat lightcurves were generated. One of these is visible in Fig. \ref{fig:ArtificialLightcurve}. These were added to an existing dataset of 864 lightcurves containing exoplanet transits.
    
    The results of the training, test and subsequent evaluations of both models are below. 
    
    \begin{table}[H]
        \centering
        \begin{tabular}{ccc}
             &  SVM& CNN\\
             Train Time&  2.43 $\pm$ .0429 s & 204.12 s\\
             Predict Time&  0.88 $\pm$ .0884 s &  2.86 s\\
             Accuracy&  93.98 $\pm$ 4.65\% & 98.34\%\\
             F1 Score& 0.9689 $\pm$ .0215 & 0.99\\
        \end{tabular}
        \label{tab:my_label}
    \end{table}

    The results of these tests indicate that, although the SVM achieved approximately 4.4\% lower accuracy and an F1 score approximately 0.02 points lower than the CNN, it significantly outperformed the CNN in terms of speed. The SVM was 84 times faster during training and 3.25 times faster in making predictions.

    These statistics demonstrate that, even with minimal hardware and human intervention, a One-Class SVM can achieve competitive levels of performance in detecting exoplanets from lightcurve data. While a CNN-based approach maintains superior accuracy and F1 scores overall, the SVM-based approach achieves accuracy within 5\% of the CNN's score while being well over an order of magnitude faster in training and over 3 times faster in predictions.
\section{Conclusions}
    With no signs of the search for exoplanets stagnating or coming to a halt, there will likely remain a need in the future for efficient, performant machine learning models to automatically detect exoplanets. With the TESS mission observing nearly twice the amount of stars that the Kepler mission did (\cite{Borucki2010} \cite{Ricker2014}), it stands to reason that the amount of data from future missions will follow this trend as optics improve and telescope sizes increase. To parse and classify this data in a reasonable amount of time, a new generation of accurate yet highly efficient machine learning models will likely be required.

    While by no means perfect, the data collected suggest that One-Class SVMs can be harnessed to perform data analysis many times faster than one of the most commonly used methods using widely available, unspecialized hardware. This research shows that One-Class SVMs can be a powerful tool in cases where making predictions on lightcurve data quickly is important. In its current form, this algorithm could have applications in large surveys of lightcurve data, false positive detection, or in other cases where performance and efficiency need to be prioritized.

    In addition, the lack of a requirement for manually-labelled datasets add to the simplicity of the One-Class SVM approach, as well as increasing speed during preprocessing.
    
    The main area found where the SVM-based approach was less effective than the CNN-based approach is accuracy. While staying within 5\% of the accuracy of a CNN, there remains room for improvement and optimization of this One-Class SVM-based method. Because of this, in scenarios where speed or efficiency are not important, a CNN-based approach might be a more preferable option. In the future, training a One-Class SVM on larger datasets, as well as taking other optimization steps, will likely yield increased accuracy while still being faster than CNN-based counterparts.
\bibliography{bibliography}
\end{document}